\documentclass[11pt]{article}

\usepackage[a4paper,margin=1in]{geometry}
\usepackage{amsmath,amssymb,amsfonts}
\usepackage{graphicx}
\usepackage{bm}
\usepackage{cite}
\usepackage{hyperref}
\usepackage{color}
\usepackage{float}

\title{Deterministic cascade coarsening in a Bistable Gene Toggle model}

\author{
Priyanka D. Bhoyar\thanks{Department of Physics, Seth Kesarimal Porwal College, Rashtrasant Tukadoji Maharaj Nagpur University, Kamptee 441001, Maharashtra, India}
\and
Prashant M. Gade\thanks{Department of Physics, RJ College of Arts, Science and Commerce, Ghatkopar West, Mumbai 400086. Maharashtra. India}
}

\date{\today}

\begin{document}

\maketitle

\begin{abstract}
We investigate deterministic coarsening dynamics in a spatially extended bistable gene toggle model with diffusive coupling. Unlike classical curvature-driven coarsening, where domain walls move continuously and annihilate gradually, the present system exhibits a qualitatively different mechanism. The domain walls remain pinned for long intervals and disappear abruptly through collective cascade events. The density of domain walls decays approximately as $\rho(t)\sim t^{-\delta}$, but the coarsening exhibits clear log-periodic oscillations superimposed on the power-law behavior.  For all values of the promoter strength $\alpha$ considered, the measured exponent satisfies $\delta<0.5$, indicating a systematic deviation from the classical Allen--Cahn prediction $\delta=1/2$ for curvature-driven coarsening. We show that log-periodic oscillations are not controlled by the density of domain walls, but by the \emph{domains that disappear} in each cascade. The average size of disappearing domains grows roughly linearly with cascade index, producing a constant geometric spacing of cascade times, consistent with discrete scale invariance.
\end{abstract}

\section{Introduction}
\label{sec1}
Gene regulatory networks frequently exhibit nonlinear dynamics that enable cells to make discrete decisions between alternative expression states. A canonical example of this behavior is the genetic toggle switch, one of the simplest and most extensively studied regulatory motifs. Regulatory switching behavior was first recognized in natural gene regulatory systems such as the lac operon \cite{Monod1961}, and later synthetically realized  by Gardner et al. \cite{Gardner2000}. The toggle switch consists of two genes that mutually repress each other, giving rise to bistability. As a result, the system can stably occupy one of two expression states, characterized by high expression of one gene and low expression of the other.

In multicellular contexts, cells do not function in isolation. They interact with neighboring cells through diffusive signaling, direct contact, and shared environmental cues \cite{Wolpert2015,Alberts2015}. Such interactions can lead to spatial correlations in gene expression, resulting in the formation of extended domains of cells in similar states \cite{Kondo2010,Gregor2007}. Similar mechanisms have been explored in both natural systems and synthetic multicellular gene circuits, where local communication between bistable cells can generate coordinated spatial patterns\cite{basu2005synthetic}. The emergence and evolution of these domains play a central role in biological processes such as tissue patterning, developmental organization, and collective cellular responses \cite{Turing1952,Wolpert1969}. Understanding how spatial gene-expression patterns form, persist, and reorganize remains a key problem at the interface of biology and physics \cite{Cross1993,Murray2002}.

Spatial coupling between cells can drive synchronization of gene-expression states, leading to a progressive reduction in phenotypic diversity as domains grow and coarsen over time. While such coarsening processes are often assumed to occur smoothly, many biological systems exhibit intermittent and burst-like dynamics, where gene-expression changes occur in discrete pulses rather than continuously~\cite{raj2008stochastic,lahav2004dynamics}. This raises the question of whether spatially coupled populations of cells can exhibit analogous intermittent dynamics at the collective level.
Motivated by these considerations, we investigate a spatially extended bistable gene toggle model with diffusive coupling. We show that domain coarsening in this system proceeds through discrete cascade events, in contrast to conventional curvature-driven coarsening \cite{Bray1994,Allen1979}. This mechanism gives rise to log-periodic oscillations superimposed on the power-law decay of domain walls, indicating the presence of discrete scale invariance in the dynamics. Our results suggest that deterministic intercellular interactions can generate both spatial synchronization and temporally organized switching events, providing a possible mechanism for coordinated yet intermittent gene-expression dynamics in multicellular systems.

Bistable gene regulatory circuits
are commonly modeled using nonlinear differential equations, where alternative expression states correspond to distinct attractors of the dynamics \cite{Ferrell2012}. Mathematically, a symmetric genetic toggle switch can be
described by:

\begin{align}
\frac{dA}{dt} = \frac{\alpha}{1 + B^n} - \gamma A, 
\end{align}
\begin{align}
\frac{dB}{dt} = \frac{\alpha}{1 + A^n} - \gamma B,
\end{align}
where $A$ and $B$ denote the concentrations of the two 
repressor proteins, $\alpha$ is the promoter strength (maximum 
production rate), $\gamma$ is the degradation rate, and $n$ is
the Hill coefficient, representing the cooperativity of 
repression.

The nonlinear Hill-type repression provides the feedback 
necessary for bistability. 
For sufficiently strong cooperativity and promoter
strength the system exhibits two stable 
steady states: a high-$A$/low-$B$ state and a low-$A$/high-$B$
state, separated by an unstable intermediate equilibrium, consistent with earlier theoretical analyses of mutually inhibitory gene networks \cite{Gardner2000,Cherry2000}.

When spatial interactions between cells are incorporated, these models naturally extend to reaction–diffusion or lattice-based frameworks that can generate collective dynamics and spatial pattern formation through the interplay of local bistable kinetics and diffusion \cite{Cross1993,Murray2002}. In such systems, interfaces separating neighboring domains are governed by the balance between reaction and diffusion processes and are often described by equations analogous to the Allen–Cahn equation \cite{Allen1979}. Fundamental analyzes of reacting and diffusing media have established the existence and stability of propagating fronts and their role in organizing spatial patterns \cite{Fife1979}.

When multiple stable states are present, spatial domains evolve through a process known as phase ordering. During this process, smaller domains shrink while larger domains expand, leading to a gradual increase in the characteristic domain size. According to Bray’s theory of coarsening, the average domain size grows as $L(t) \sim t^{1/2}$ for non-conserved order parameters, a scaling law that has been extensively validated in bistable reaction–diffusion systems \cite{Bray1994}.

In this work, we investigate deterministic coarsening dynamics in a one-dimensional array of diffusively coupled bistable gene toggle switches. Unlike classical curvature-driven phase ordering, domain walls remain nearly stationary for long periods and disappear abruptly through collective cascade events. The resulting dynamics exhibit power-law decay of domain walls with superimposed log-periodic oscillations, indicating discrete scale invariance \cite{sornette1998discrete}. We show that the evolution is controlled by the population of  domains disappearing in each cascade rather than by the total domain wall density. Finally, we analyze how promoter strength $\alpha$ and coupling strength D govern bistability, interface stability, cascade statistics, and collective ordering.

\section{Model and Simulation}
While the classical toggle describes intracellular dynamics within a single cell, many biological systems involve spatial interactions between neighboring cells. To investigate collective dynamics arising from intercellular coupling, we extend the toggle switch to a one-dimensional lattice of diffusively coupled bistable units with periodic boundary conditions. The spatially extended model is given by:

\begin{align}
\frac{dA_i}{dt} =
\frac{\alpha_1}{1 + B_i^n}- \gamma A_i+ D(A_{i+1} + A_{i-1} - 2A_i),
\end{align}

\begin{equation}
\frac{dB_i}{dt} =
\frac{\alpha_2}{1 + A_i^n}
- \gamma B_i,
\end{equation}

where $i = 1,2,\dots,N$. Here, $A_i$ and $B_i$ denote the concentrations of the two repressor proteins at site $i$, $\alpha_1$ and $\alpha_2$ represent the promoter strengths (maximum production rates) for the two genes, $\gamma$ is the degradation rate, $n$ is the Hill coefficient describing the cooperativity of repression, and $D$ denotes the diffusive coupling strength between neighboring sites. Numerical simulations were performed on a lattice of size N~=50000 with periodic boundary conditions. The equations were integrated using a fourth-order Runge–Kutta method with timestep $\Delta$ t = 0.01 up to $t_{\max} \sim 2.5\times10^6$.

Since the two molecular species need not possess identical mobilities, we study the minimal case in which A diffuses and B remains local. Such unequal effective diffusivities are common in intracellular signalling systems due to binding and localization effects\cite{kholodenko2009spatially}. Diffusive coupling between neighboring sites can then generate collective dynamics and spatial pattern formation \cite{Cross1993}.
Since each cell is bistable, spatial interactions can lead to the
formation of domains in which neighboring cells share the same
expression state. The boundaries separating such regions act as
domain walls (or fronts), whose motion determines the evolution
and eventual coarsening of spatial patterns.

Initial concentrations ($A_i$) and ($B_i$) were drawn independently from a uniform random distribution on the interval $[0,1]$. The local state of each lattice site is determined by comparing the
concentrations of the two genes.  A site is classified as $A$
dominated if $A_i>B_i$ and $B$-dominated otherwise, and we define an effective spin variable in the following manner:

\begin{align*}
s_i=
\begin{cases}
+1 & \text{if } A_i > B_i \\
-1 & \text{if } A_i < B_i .
\end{cases}
\end{align*}

A domain wall is defined as
a transition between neighboring sites with different states.
Thus, an interface occurs whenever $s_i \neq s_{i+1}$. The
density of domain walls $\rho(t)$ is obtained by
counting such transitions and normalizing by the system size.
Spatial domains correspond to contiguous regions of identical
spin values. The results shown are obtained by averaging over atleast 10 realizations. 

\subsection{Qualitative Analysis of the Reaction Term}

The deterministic coarsening dynamics can be understood qualitatively by examining the single cell reaction term. Defining
\begin{align}
f(A) = \frac{\alpha_1}{1 + B(A)^n} - \gamma A, \quad    
\end{align}
\begin{align}
B(A) = \frac{\alpha_2}{\gamma (1 + A^m)},
\end{align}

To understand the deterministic coarsening dynamics, we examine the
single cell reaction term after eliminating $B$ via its steady state
expression. For $m=n=2$ and $\alpha_1=\alpha_2=10$, we have
\begin{align*}
f(A)
&= \frac{10}{1 + B(A)^2}-\gamma A\\
&=\frac{10}{1 + \dfrac{100}{\gamma^2 (1 + A^2)^2}}-\gamma A.\\
&=\frac{10 \gamma^2 (1 + A^2)^2}{\gamma^2 (1 + A^2)^2 + 100}-\gamma A. 
\end{align*}

Steady states satisfy $f(A)=0$, which implies\\
\begin{align*}
10 \gamma^2 (1 + A^2)^2 =
\gamma A \left[\gamma^2 (1 + A^2)^2 + 100 \right].
\end{align*}

\begin{figure}[h!]
\centering
\includegraphics[width=0.8\linewidth]{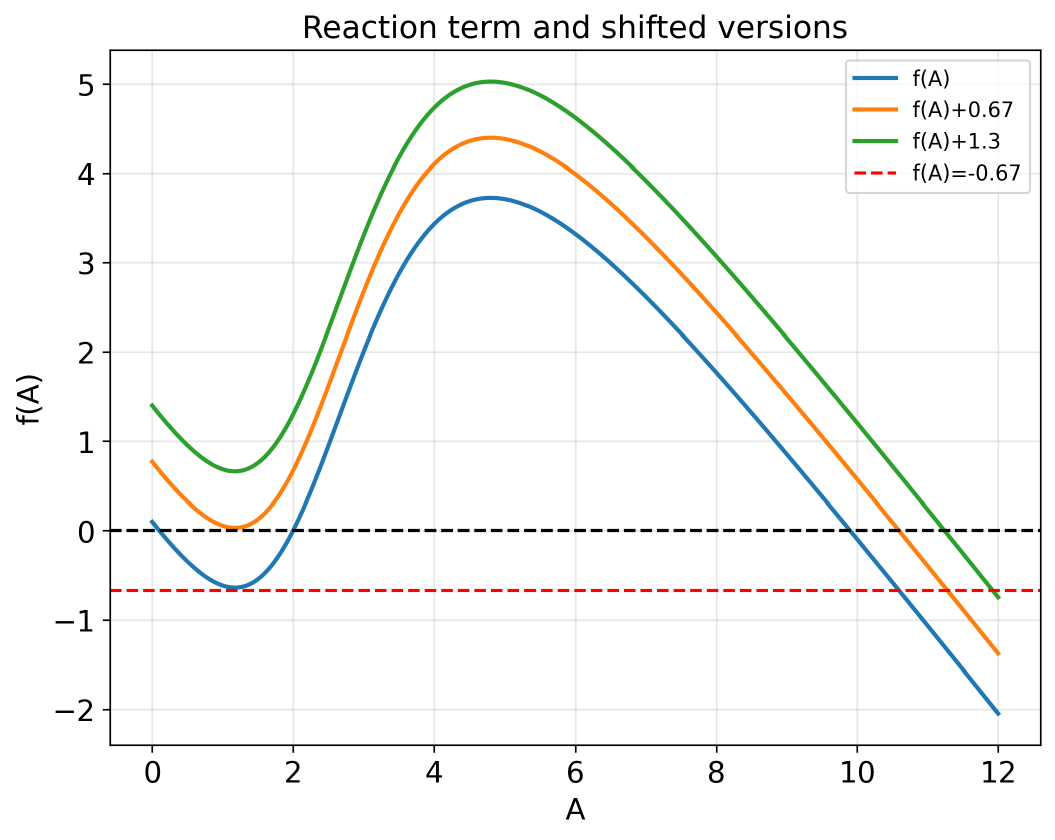}
\caption{
Qualitative shape of the reaction term $f(A)$ (solid line) and shifted versions $f(A)+0.67$ and $f(A)+1.3$. The dashed line shows $f(A)=-0.67$. The shift illustrates the disappearance of small roots corresponding to front depinning. 
}
\label{fig:fa}
\end{figure}

This is a fifth-degree polynomial equation in $A$.
Nevertheless, for the parameter ranges ($A \ge 0$), the function $f(A)$ possesses two
turning points, producing three real steady states: two stable fixed
points separated by one unstable state. This effective ``cubic''
phase-portrait topology underlies the bistability of the toggle switch. One observes that $f(A)$ has a local minimum around $A \simeq 1.15$ (see Figure.~\ref{fig:fa}). Shifting $f(A)$ by $+0.67$ effectively removes the small root, corresponding to the disappearance of the lower stable state in a single cell. This process can be interpreted as \emph{front depinning}. The cell is pushed out of the low-$A$ state due to contributions from its neighbors.

Specifically, the diffusive coupling term
\begin{align*}
D(A_{i+1} + A_{i-1} - 2 A_i)
\end{align*}
provides the necessary bias. For instance, if $A_i$ and $A_{i+1}$ are in the small-$A$ state ($A\sim 1.15$) while $A_{i-1}$  is large ($A\sim 8.5)$, the Laplacian term is positive, promoting an increase in $A_i$ and thereby triggering depinning at $D=0.091$.

An exact analytic calculation of the depinning threshold is challenging because the system is not strictly one-dimensional. $A_i$ does not remain at very low levels for long, so the effective $D$ required for decoupling is slightly higher than predicted by a naive one-dimensional estimate.

\begin{figure}[h!]
\centering
\includegraphics[width=1.0\linewidth]{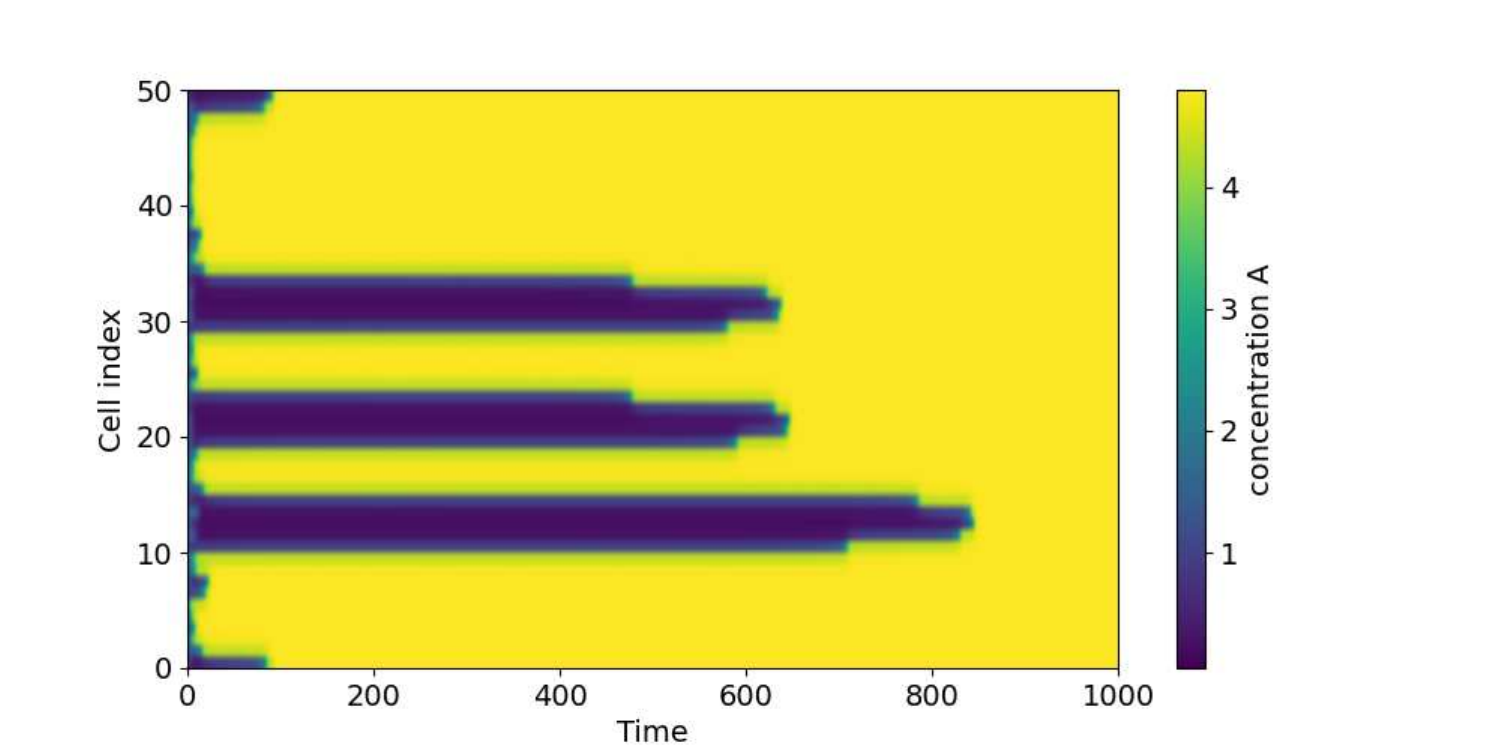}
\caption{Spatiotemporal evolution of concentration $A(x,t)$. Domain walls remain stationary and disappear abruptly.}
\label{fig:domain}
\end{figure}

\section{Spatiotemporal Dynamics}
Figure.~\ref{fig:domain} shows the spatiotemporal evolution of 
the concentration \(A_i(t)\) in a one–dimensional array of
bistable gene toggle switches with diffusive coupling strength
\(D = 0.116808889\) and \(\alpha = 5\). In the color map, the x-axis corresponds to time, 
while the y-axis represents the discrete cell index \(i\).
The color encodes the instantaneous value of \(A_i(t)\).
Dark (low) values of \(A_i\) correspond to one stable
gene–expression state and bright (high) values corresponds to the
alternative stable state, so each horizontal stripe represents
the temporal evolution of a single cell in the array.

Extended regions of similar color indicate spatial domains in
which neighboring cells share the same gene-expression state.
These domains persist for long times and are separated by
sharp interfaces (domain walls). Domain annihilation occurs
in discrete, step-like events where entire domains collapse
suddenly rather than gradually. As a result, the number of
interfaces decreases in abrupt jumps. Unlike conventional
diffusive coarsening, the domain walls remain nearly
stationary for long periods between these annihilation
events, indicating intermittent dynamics in the coarsening
process.

\section{Temporal Decay of Domain Walls}
\begin{figure}[h!]
\centering
\includegraphics[width=0.8\linewidth]{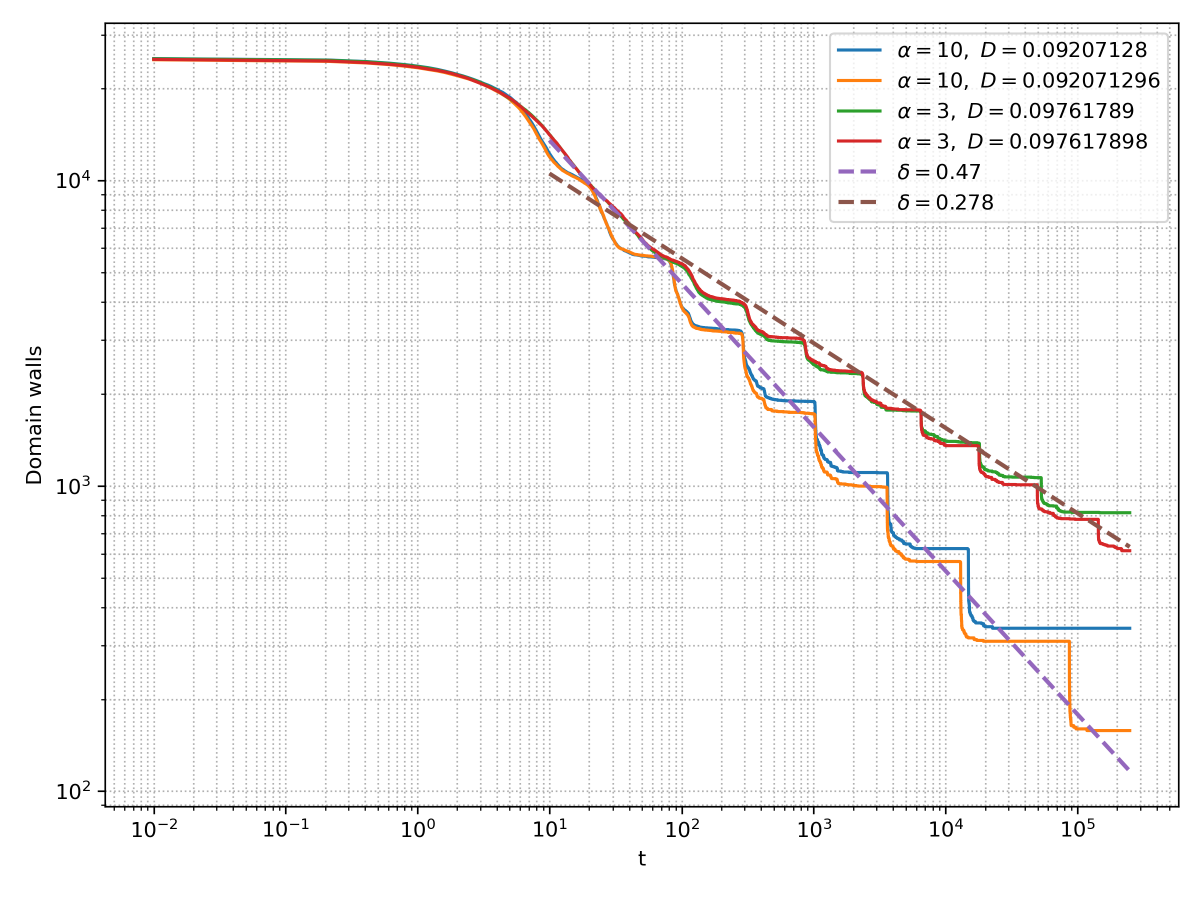}
\caption{Unnormalized domain-wall density $\rho(t)$ as a function of time on logarithmic scales for $\alpha=3$ and $\alpha=10$ at diffusion coefficients close to the depinning threshold. Dashed lines indicate the reference power laws $t^{-0.278}$ and $t^{-0.47}$.}
\label{fig:decay}
\end{figure}

The domain density $\rho(t)$
decays approximately as
\begin{align*}
\rho(t) \sim t^{-\delta},
\end{align*}
Figure~\ref{fig:decay} compares the coarsening dynamics for $\alpha=3$ and $\alpha=10$ near the depinning threshold. In both cases, the domain-wall density decays algebraically, $\rho(t)\sim t^{-\delta}$, but with markedly different exponents. While the $\alpha=10$ case exhibits a decay close to the classical Allen--Cahn value $\delta=1/2$, the $\alpha=3$ case displays significantly slower coarsening, characterized by an effective exponent $\delta\approx0.278$. The dashed lines in Fig.~\ref{fig:decay} show the corresponding reference power laws.

The geometric spacing of cascade times is directly
related to the frequency of the
log-periodic oscillations for a log-periodic
correction of the form:
\begin{align*}
\rho(t) \sim t^{-\delta}\left[c_0 + c_1 \cos(\omega \log t + \phi)\right],
\end{align*}

\begin{figure*}[t!]
    \centering
       \includegraphics[width=0.48\textwidth]{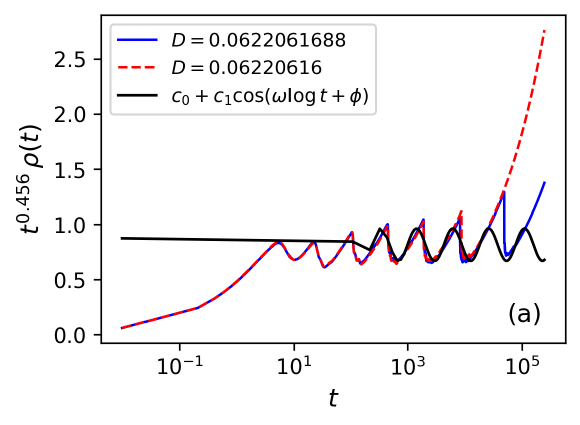}
       \includegraphics[width=0.48\textwidth]{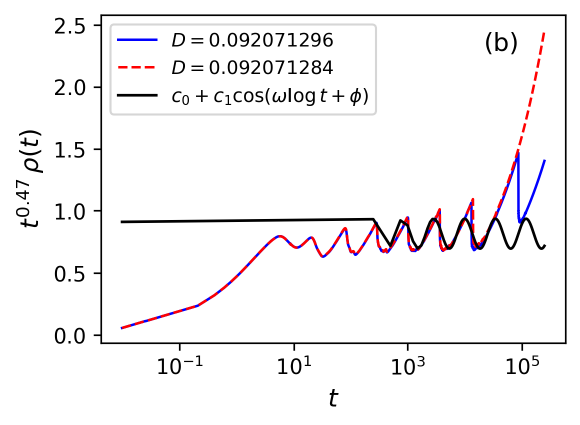}
    \includegraphics[width=0.48\textwidth]{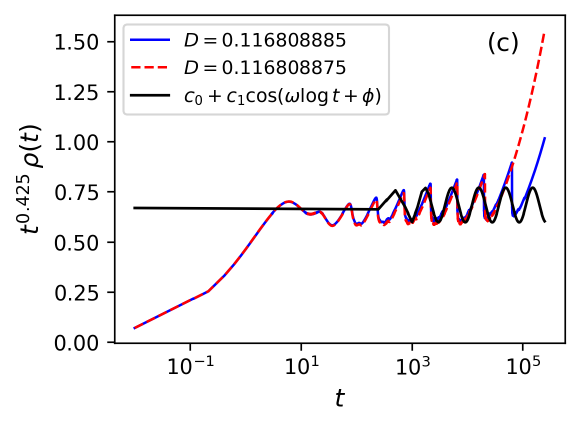}
      \includegraphics[width=0.48\textwidth]{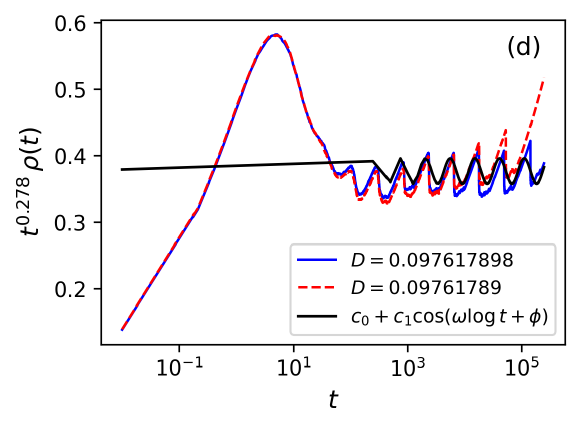}
    \caption{
 Rescaled domain density $t^{\delta}\rho(t)$ versus $\log t$ for (a) $\alpha=20$ ($D_c=0.0622061688$), together with the fitted form $0.81680-0.14642\cos(4.33166\log t+15.40398)$ and $\delta=0.456$; (b) $\alpha=10$ ($D_c=0.092071296$), together with the fitted form $0.81871+0.12053\cos(5.05515\log t+10.05278)$ and $\delta=0.47$; (c) $\alpha=5$ ($D_c=0.116808885$), together with the fitted form $0.68368-0.08534\cos(5.59412\log t+5.53014)$ and $\delta=0.425$; and (d) $\alpha=3$ ($D_c=0.097617898$), together with the fitted form $0.37696+0.01890\cos(6.12832\log t-3.93890)$ and $\delta=0.278$. The parameters are obtained from nonlinear least-squares fits to the log-periodic form $c_0+c_1\cos(\omega\log t+\phi)$.}
    \label{rescaled}
\end{figure*}

The parameters were obtained using non-linear
least-squares fitting. The uncertainties of the fitted 
parameters were estimated from the
covariance matrix returned by the fitting procedure. 
The rescaled domain density $t^{\delta} \rho(t)$ vs time (log scale) for $\alpha_1=\alpha_2=20,10,5,3$ is plotted in Figure.\ref{rescaled} along with the fitting curve.
The parameters obtained from nonlinear fits of the
rescaled domain density $t^{\delta}\rho(t)$ for different values of the control parameter $\alpha$ are tabulated in Table.\ref{table1}.

A systematic trend is observed as the promoter strength $\alpha$ is reduced. There is monotonic increase of the log-periodic frequency $\omega$ from approximately 4.33 for $\alpha=20$ to 6.13 for $\alpha=3$. Equivalently, the associated temporal scaling ratio $\lambda=\exp(2\pi/\omega)$
 decreases from about 4.13 to 2.79 leading to a denser hierarchy of characteristic timescales. The product $\delta \omega$ is approximately 2 for all values of $\alpha$ studied, suggesting a possible inverse scaling $\delta \sim 1/\omega$. Confirming this relation would require simulations over a wider parameter range.

An even more surprising feature of the dynamics is the behavior of the coarsening exponent $\delta$. For all values of the promoter strength $\alpha$ considered, the measured exponent remains below the classical Allen--Cahn value $\delta=1/2$ expected for non-conserved order-parameter dynamics\cite{Allen1979}. Furthermore, $\delta$ decreases systematically with decreasing $\alpha$, reaching $\delta \simeq 0.28$ for $\alpha=3$.

This value is significantly smaller not only than the Allen--Cahn exponent but also than the Lifshitz--Slyozov/Cahn--Hilliard exponent $\delta=1/3$ commonly associated with conserved dynamics\cite{Bray1994}. The observation is particularly striking because the present model possesses neither a conserved order parameter nor the curvature-driven interface motion that underlies these classical coarsening theories.

A possible qualitative explanation is that decreasing $\alpha$ broadens the domain walls and extends the range of front--front interactions. It remains unclear how these effects quantitatively determine the observed power-law exponent. We are not aware of any  theoretical framework predicts that can explain  the values of  $\delta$ observed here, particularly the exceptionally small value $\delta \simeq 0.28$ obtained for $\alpha=3$. Understanding the origin of this anomalously slow coarsening therefore remains an interesting open problem.





\begin{table}[h]
\centering
\begin{tabular}{c c c c c c  }
\hline
$\alpha$ & $\delta$ & $D_c$& $c_0$ & $c_1$ & $\omega$  \\
\hline
20 & 0.456 & 0.0622061688 & 0.81680   & -0.14642   & 4.33166$\pm$0.00170  \\

10 & 0.47 & 0.092071296 & 0.81871   & 0.12053   & 5.05515$\pm$0.00171  \\

5  & 0.425 & 0.116808885 & 0.68368 & -0.08534 & 5.59412$\pm$0.00081 \\

3  & 0.278 & 0.097617898 & 0.37696 & 0.01890 & 6.12832$\pm$0.00097  \\
\hline
\end{tabular}
\caption { Fitted parameters of the log-periodic form $c_0 + c_1 \cos(\omega \log t + \phi)$ obtained from nonlinear fits to the rescaled domain density $t^{\delta}\rho(t)$ for different values of the control parameter $\alpha$. $\phi$ is also fitted with error bar in second decimal but not shown.}
\label{table1}
\end{table}

On the other hand, the decrease in $\lambda$ with decreasing  $\alpha$ can be qualitatively understood. As $\alpha$ decreases, the effective reaction function $f(A)$ becomes flatter near the low stable state $A_L$, leading to a reduction in the magnitude of $|f'(A_L)|$. Since the front localization parameter scales as $\kappa = \sqrt{-f'(A_L)/D}$, this implies a decrease in $\kappa$, resulting in more weakly localized fronts with longer exponential tails. Consequently, front-front interactions decay more slowly with domain size, enhancing the coupling between interfaces over larger distances \cite{rubinstein1993front,van2010front}. Domains  become unstable at smaller characteristic sizes, causing successive collapse events to occur closer together on a logarithmic time scale. In the cascade picture, this leads to shorter geometric spacing between successive collapse events, reflected in a decrease of the scaling ratio $\lambda = e^{\kappa}$. Thus, the observed decrease in $\lambda$ with decreasing $\alpha$ is qualitatively consistent with the reduction of $|f'(A_L)|$, which governs front localization and interaction strength in the system. 

Discrete scale invariance requires invariance
under the transformation 
$t \to \lambda t$, which implies
$\omega \log \lambda = 2\pi.$
Hence, the temporal scaling ratio is
$\lambda = e^{2\pi/\omega}$. For example, with $\omega=5.05515$ we obtain
$\lambda = \exp\!\left(\frac{2\pi}{5.05515}\right) \approx 3.465.$ (For $\alpha$=20,~5,~3 $\lambda$=4.128,~3.074,~2.787 respectively). 
The observed geometric cascade spacing should be
quantitatively consistent with the
measured log-periodic frequency, confirming that both
originate from the same
underlying discrete scale invariance. We will observe
in next section, the ratio is
indeed close to the expected value.

\section{Disappearing domains and cascade timescales}
\label{secn1}
Let $L_i$ denote the length of an \emph{active domain} that disappears in the $i^{th}$ cascade. The collapse of such domains is governed by front interactions:
\begin{equation}
\frac{dL_i}{dt} = -v_0 \, e^{-\kappa L_i}
\end{equation}
where $v_0$ is a kinetic prefactor and $\kappa$ is the front localization parameter determined by the local bistable dynamics. 

\begin{figure}
\includegraphics[width=.9\linewidth]{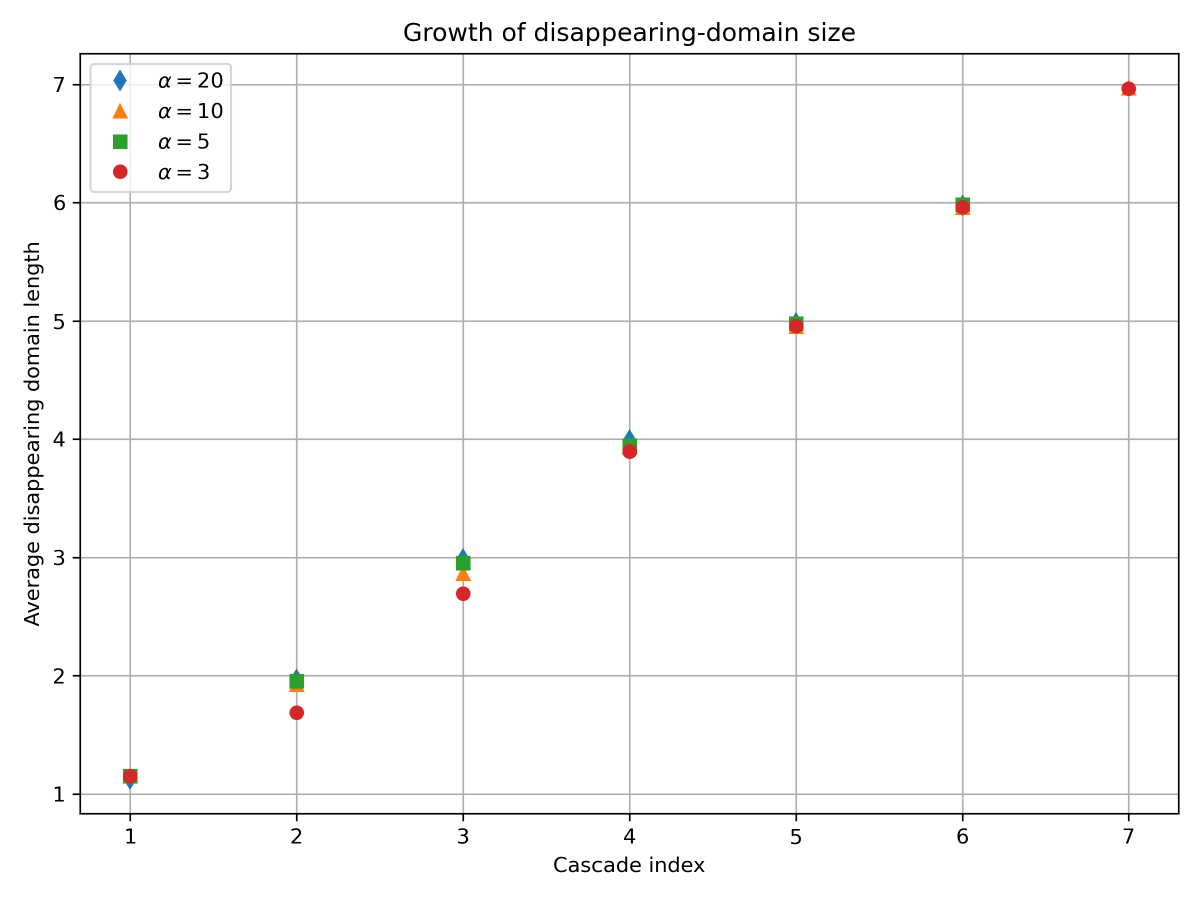}
\caption{
Average disappearing-domain length $\langle L_{\rm dis}\rangle$ versus cascade index for $\alpha=3,5,10$, and $20$ at their respective critical coupling strengths.
}
\label{fig:Ldis}
\end{figure}

To identify the domains responsible for cascade events, we measured the size of domains that disappear between successive measurement times chosen within the flat plateau regions separating two cascade events. At these times the domain configuration is essentially stationary, allowing a clear comparison of the states before and after a cascade. An effective spin variable was assigned to every lattice site.
The spin configuration at a given measurement time was compared with that at the next measurement time. A disappearing domain was defined as a contiguous domain present in the earlier configuration whose sites had completely switched state in the later configuration, indicating that the domain had vanished during the intervening cascade. The length $L_{\rm dis}$ of a disappearing domain was taken as the number of lattice sites contained within that domain. Averaging over all disappearing domains yielded the mean disappearing-domain size $\langle L_{\rm dis}\rangle$. Since the collapse of a domain is governed by the interaction between the two fronts bounding it, $L_{\rm dis}$ directly measures the front separation appearing in the front--interaction law
$
t_c(L)\sim e^{\kappa L},
$
and therefore provides the natural length scale controlling the cascade times. To test the proposed cascade mechanism, we measured the average disappearing-domain length $\langle L_{\rm dis}\rangle$ as a function of cascade index. As shown in Fig.~\ref{fig:Ldis}, the characteristic size of the domains that disappear in successive cascades increases approximately linearly with cascade index for all values of $\alpha$ studied. The slope is close to one lattice site per cascade, indicating that each successive cascade is typically dominated by domains whose size is approximately one lattice site larger than those involved in the previous cascade. Since the collapse time grows exponentially with domain size,
The collapse time of the $i^{th}$ domain:
\begin{equation}
t_c^{(i)} \sim \frac{e^{\kappa L_i}}{\kappa v_0 }.
\end{equation}

Crucially, the density of domain walls is not the controlling factor. Instead, the \emph{average length of disappearing domains} $\langle L_i \rangle$ increases roughly linearly with cascade index $i$, leading to a constant geometric spacing:
\begin{equation}
\lambda_t = \frac{t_c^{i+1}}{t_c^{i}} \simeq 3.476.
\end{equation}
Since, $t_c(L+1)=e^{\kappa}t_c(L)$
successive cascade times are expected to be geometrically spaced whenever the characteristic disappearing domain length increases by approximately one lattice site between cascades. 
 Because domain lengths are discrete lattice variables, the characteristic disappearing domain size can change only in integer increments. This ratio is observed for $\alpha=10$. While this feature is seen for other values of 
 $\alpha$, the precise value of the ratio  changes with the parameter.

Thus, the log-periodic oscillations in $\rho(t)$ emerge naturally from the discrete hierarchy of disappearing domain lengths. The associated logarithmic period is determined by the geometric spacing of cascade times generated by domain size quantization, rather than by the density of domain walls.

\section{Connection to Discrete Scale Invariance}

A striking feature of the deterministic cascade coarsening observed in the bistable gene toggle model is the emergence of discrete scale invariance (DSI) in the temporal evolution of the domain wall density. 
Unlike many classical realizations of discrete scale invariance, where the hierarchy is imposed by an underlying fractal or hierarchical structure, the present model evolves on a uniform one-dimensional lattice with identical local dynamics. Here, the apparent DSI emerges dynamically from the combination of three key mechanisms:

\begin{enumerate}
    \item \textbf{Discrete domain lengths:} Domain lengths can change only in integer increments, with one lattice site representing the smallest possible change.
    
    \item \textbf{Exponential dependence of collapse times on domain length:} Each domain  that will disappear in a cascade collapses on a timescale
    \[
        t_c \sim e^{\kappa L_i},
    \]
    where $L_i$ is the domain length and $\kappa$ is set by the bistable kinetics and diffusion. Since the collapse time increases exponentially with domain size, larger domains persist for progressively longer periods. Because the characteristic disappearing domain length increases in approximately discrete unit increments, successive cascade events occur at geometrically spaced times.

    \item \textbf{Linear growth of disappearing domain lengths:} In our simulations, the average length of such domains grows approximately linearly with cascade index, $L_i \sim i$. Combined with the exponential dependence of collapse times, this leads to
    \[
        t_c^{i+1}/t_c^i \sim e^{\kappa (L_i + 1)}/e^{\kappa L_i} = e^{\kappa} = \lambda_t,
    \]
    producing a nearly constant geometric ratio between successive cascade times. In our system, this ratio is $\lambda_t \simeq 3.5$(for $\alpha=10$), consistent with the observed log-periodic oscillations.
\end{enumerate}

Thus, the resulting discrete scale invariance is therefore temporal rather than spatial. The hierarchy arises from the sequence of cascade times rather than from any underlying geometric hierarchy in the lattice itself.
In contrast to conventional DSI systems, where scale invariance is encoded in the structure of the medium, the hierarchy here is generated dynamically through the interplay of bistability, exponentially weak front interactions, and discrete domain sizes.

\section{Discussion and Conclusions}
The coarsening dynamics in the spatially extended bistable gene toggle model differ from those of conventional bistable systems. In classical phase ordering, domain walls move continuously and domains disappear gradually through interface annihilation. In contrast, our system exhibits long periods of little change interrupted by sudden cascade events in which entire  domains vanish. As a result, the evolution is controlled not by the density of domain walls, but by the  domains that disappear during successive cascades.

A key feature of the system is the critical coupling strength $D_c$, which separates pinned and mobile front dynamics. For $D<D_c$, domain walls remain pinned, leading to a frozen pattern of stable domains. For $D>D_c$, interfaces become mobile, and  the system rapidly approaches a homogeneous state. Near $D_c$, front interactions give rise to cascade events that dominate the long time evolution. Thus, the critical coupling marks the boundary between frozen, critical, and rapidly coarsening regimes.

The emergence of discrete scale invariance follows naturally from the cascade dynamics. Disappearing domain lengths are discrete lattice variables, while the collapse time of a domain increases exponentially with its size. Together, these features generate geometrically spaced cascade times and give rise to the observed log-periodic oscillations in the domain wall density. Unlike conventional examples of discrete scale invariance, where hierarchy is built into a fractal or hierarchical structure, the hierarchy here emerges dynamically from deterministic front interactions on a uniform lattice.

From a biological perspective, the present results suggest that spatially coupled bistable gene networks need not evolve smoothly toward a uniform state. Instead, collective gene-expression patterns may reorganize through intermittent cascade-like events, leading to periods of apparent stability punctuated by rapid transitions. Such behavior could provide a mechanism for coordinated yet episodic changes in multicellular systems.

In summary, we have investigated deterministic coarsening in a one-dimensional lattice of diffusively coupled bistable gene toggle switches. The domain wall density decays approximately as $\rho(t)\sim t^{-\delta}$ with superposed log-periodic oscillations. For all values of the promoter strength $\alpha$ considered, the measured exponent satisfies $\delta<0.5$, indicating a systematic deviation from the classical Allen--Cahn prediction $\delta=1/2$ for curvature-driven coarsening. We show that these oscillations arise from the dynamics of  domains, whose collapse times increase exponentially with domain length. As the characteristic disappearing domain length grows approximately linearly with cascade index, successive cascade events become geometrically spaced in time, yielding a scaling ratio $\lambda_t \simeq 3.5$ (for $\alpha=10$). These results establish a direct connection between front interactions, cascade coarsening, and temporal discrete scale invariance in a deterministic bistable system. More broadly, they demonstrate how bistability, diffusion, and lattice discreteness can generate hierarchical temporal organization without any underlying fractal or hierarchical spatial structure.

An interesting possibility is that the deterministic cascade coarsening observed here is not unique to the bistable gene-toggle model. Similar qualitative features have been reported in other deterministic spatially extended systems. For example, at the $T_3$ synchronization transition in coupled $q$-deformed logistic maps\cite{sabe2024synchronization}, the order parameter exhibits power-law decay with visible log-periodic modulations, while the spatiotemporal dynamics are dominated by localized defects that remain nearly stationary and disappear over time rather than spreading diffusively (See Fig. 7 and Fig. 3c in \cite{sabe2024synchronization}). Although only a few oscillation cycles were visible in that system, preventing a quantitative determination of the associated discrete scale invariance, the dynamics share several qualitative features with those reported here. It is therefore plausible that temporal discrete scale invariance associated with the annihilation of localized structures may occur in a broader class of deterministic spatially extended systems. One reason such behavior may not have been widely recognized is that its numerical detection is particularly demanding. Resolving log-periodic corrections requires simulations over many decades in time, sufficiently large system sizes to suppress finite-size effects, and careful tuning to the critical point. In the present study, a clear hierarchy of cascade events became visible only after extensive simulations on large lattices over very long time intervals. Similar phenomena may therefore have remained overlooked in other systems because their reliable identification requires extensive computations, large system sizes, long simulation times, and careful tuning to criticality.

Previous studies have shown that log-periodic oscillations typically arise either from intrinsic hierarchical or fractal structures \cite{sornette1998discrete,saleur1996discrete} or from disorder-induced fragmentation that generates a hierarchy of characteristic timescales \cite{bhoyar2020, bhoyar2021emergence}.Log-periodic oscillations have also been reported during the approach to material rupture and earthquakes, where hierarchical damage accumulation and crack interactions generate discrete scale invariance near the critical failure point \cite{anifrani1995universal,huang1998precursors,sornette1998discrete}.
Similar log-periodic behavior has been observed in speculative financial bubbles, where collective herding and positive feedback among market participants produce accelerating oscillations preceding market crashes \cite{johansenpredicting,johansen2000crashes,sornette1998discrete}.
By contrast, our model contains neither quenched disorder,  underlying hierarchical geometry, damage accumulation on heterogeneous substrate nor agent based feedback. Instead, the observed temporal discrete scale invariance emerges spontaneously from deterministic front interactions, discrete domain lengths, and cascade dynamics on a homogeneous lattice.
Although the present study is formulated in terms of a bistable gene-toggle network, the cascade mechanism relies only on bistability, spatial coupling, front pinning, and lattice discreteness. These ingredients occur in a broad range of biological, chemical, ecological, and physical systems. We therefore expect the disappearing-domain description and the resulting temporal discrete scale invariance to extend beyond gene-regulatory dynamics.

\appendix
\section{Appendix: Derivation of the Front Interaction Collapse Law}

In this appendix we provide a heuristic argument that the domain collapse law
$
\frac{dL}{dt} = -v_0 e^{-\kappa L}
$
used in Eq.~(7) follows naturally from the interaction between two
stationary fronts in a bistable reaction--diffusion system.

\subsection{Single Front Solution}

Consider the reaction--diffusion equation obtained after eliminating
$B$ using its quasi-steady expression,
\begin{align*}
\partial_t A = D \partial_x^2 A + f(A).
\end{align*}

A front connecting the two stable states $A_L$ and $A_H$ satisfies the
traveling-wave equation
\begin{align*}
D A'' + c A' + f(A) = 0 ,
\end{align*}
where $c$ is the front velocity and primes denote derivatives with
respect to the comoving coordinate $\xi = x - ct$.
At the critical diffusion $D_c$ the front is pinned, so $c=0$ and the
equation reduces to
\begin{align*}
D A'' + f(A) = 0 .
\end{align*}

\subsection{Exponential Tail of the Front}

To determine the asymptotic behavior of the front, we linearize around
the low stable state $A_L$:
$
f(A) \approx f'(A_L)(A-A_L).
$ Defining $\delta A = A-A_L$ gives
$
D \delta A'' + f'(A_L)\delta A = 0 .
$

Since $f'(A_L) < 0$, the solution decays exponentially,
$
\delta A(\xi) \sim e^{-\kappa |\xi|},
$
where
$
\kappa = \sqrt{-\frac{f'(A_L)}{D}} .
$
Thus perturbations generated by a front decay exponentially away from
the interface.

\subsection{Interaction of Two Fronts}

Consider now a finite domain of length $L$ bounded by two such fronts.
Because the tails decay exponentially, the perturbation generated by
one front at the position of the other scales as
\begin{align*}
\delta A \sim e^{-\kappa L}.
\end{align*}

An isolated front is stationary because the reaction and diffusion
terms balance exactly. The perturbation from the opposite front
breaks this balance and produces a small driving force of order
\begin{align*}
\Delta f \sim f'(A_L) e^{-\kappa L}.
\end{align*}

\subsection{Front Velocity}

For small driving forces, the front velocity responds linearly,
\begin{align*}
c \propto \Delta f .
\end{align*}

Hence the velocity induced by the interaction between the two fronts
scales as
\begin{align*}
c \sim e^{-\kappa L}.
\end{align*}

Since a domain is bounded by two fronts moving toward each other, the
domain length evolves according to
\begin{align*}
\frac{dL}{dt} = -v_0 e^{-\kappa L},
\end{align*}
where $v_0$ depends on the front profile and the kinetic parameters.

We note that the linearization of the reaction term around the single cell low stable fixed point $A_L$ and the resulting front tail decay rate $\kappa = \sqrt{-f'(A_L)/D}$ are only qualitative. At domain boundaries, the local concentration A differs from the isolated fixed point value $A_L$ due to coupling with neighboring cells. Thus, $\kappa$ does not provide a quantitative prediction of cascade times, but it illustrates that front interactions decay exponentially with domain size, motivating the exponential collapse law $\frac{dL}{dt} = -v_0 e^{-\kappa L}$ used in our analysis.

\section*{AI-Assisted Writing Statement}
Generative AI tools were used only to improve the language and 
clarity of this manuscript. The authors carefully reviewed and 
edited the output and take full responsibility for the final content.

\section*{Acknowledgements}
PMG thanks IMSc, Chennai, for hosting a visit and Prof. Sitabhra Sinha for discussions. PDB thanks Rashtrasant Tukadoji Maharaj 
Nagpur University for providing financial assistance (RTMNU/RDC/2024/242)


\bibliographystyle{unsrt}
\bibliography{arxiv}

\end{document}